\documentclass[a4paper,10pt]{article}
\usepackage{epsfig,epsf}
\usepackage{mathrsfs}
\usepackage{graphicx}
\usepackage{caption}
\usepackage{latexsym}
\usepackage{amsmath}
\usepackage{amssymb}
\usepackage{textcomp}
\usepackage{amsbsy}
\usepackage{tabularx}
\usepackage{array}
\usepackage{amsmath,amsfonts,amssymb}
\usepackage{hyperref}
\usepackage{multicol}
\DeclareGraphicsExtensions{.png,.eps,.pdf}
\usepackage{color}
 \usepackage[normalem]{ulem}
 
\definecolor{darkgreen}{cmyk}{1,0,1,0.4}

\long\def\/*#1*/{}
\renewcommand{\baselinestretch}{1.5}

\def\beq{\begin{equation}}
\def\eeq{\end{equation}}
\def\barr{\begin{array}}
\def\earr{\end{array}}
\def\dis{\displaystyle}
\def\tev{\, {\rm TeV}}

\def\lapp{\mathrel{\rlap{\raise.5ex\hbox{$<$}}
                    {\lower.5ex\hbox{$\sim$}}}}
\def\gapp{\mathrel{\rlap{\raise.5ex\hbox{$>$}}
                    {\lower.5ex\hbox{$\sim$}}}}

\def\gym{g_{\mbox{\lower.25ex\hbox{\tiny YM}}}}
\begin{document}

\begin{center} 
{\large \bf Stabilization of moduli in spacetime with nested warping and the UED } \\
\vspace*{1cm} 
{Mathew Thomas Arun} and {Debajyoti Choudhury} 
\\ 
\vspace{10pt} 
{\it Department of Physics and Astrophysics, University of
Delhi, Delhi 110 007, India}

\normalsize 
\end{center}
\begin{abstract}
The absence, so far, of any graviton signatures at the LHC
  imposes severe constraints on the Randall-Sundrum scenario.
  Although a generalization to higher dimensions with nested warpings
  has been shown to avoid these constraints, apart from
    incorporating several other
  phenomenologically interesting features, moduli stabilization
  in such models has been an open question.  We demonstrate here how
  both the moduli involved can be stabilized, employing slightly
  different mechanisms for the two branches of the theory. This also 
offers a dynamical mechanism to generate and stabilise
the scale for the Universal Extra Dimensions, another long-standing issue.
\end{abstract}

\newpage
\section{Introduction}
The discovery of the Higgs boson~\cite{atlas-higgs, cms-higgs}, while
seemingly completing the jigsaw that the Standard Model (SM) is, has
also brought into sharp focus a long-standing puzzle that has plagued
the SM. The very lightness of its being militates against the
conventional wisdom that the mass of a fundamental scalar should flow,
at the very least, to the next-higher scale in the theory. Several
``resolutions'' of this {\em hierarchy problem} have been proposed,
some technically natural and others not so, most of these relying on
some new dynamics and/or new states appearing at the few--TeV scale
that would serve to nullify the largest of the quantum corrections
accruing from within the SM. The continuing absence of any direct
evidence of such states, though, bring into question many such
explanations.

A particularly elegant resolution is proffered by higher-dimensional
theories. While models with large extra dimensions~\cite{Antoniadis:1998ig,ArkaniHamed:1998rs} have been quite popular, these fail to
  truly solve this problem in that these proffer no mechanism to
  stabilize the corresponding moduli. Similar is the case with
  Universal Extra Dimensions (UED)~\cite{Appelquist:2000nn} which, while
  proffering interesting phenomenological consequences, such as an
  origin of Dark Matter, flavour physics as well as collider
  signatures, again do not really solve the problem of large
  hierarchies. Quite the opposite is the case of theories with a
warped geometry~\cite{Gogberashvili:1998vx, Gogberashvili:1998iu,
  Randall:1999ee}, wherein one assumes space-time to be a slice of
AdS$_5$, bounded by two 3-branes, on one of which (the TeV brane) the
SM fields are confined. There is but one fundamental scale (the scale
of gravity $M_5$, very close to the derived scale $M_{\rm Planck}$) in
the theory, and the smallness of the electroweak scale (with respect
to $M_5$) is only an apparent one, caused by the non-trivial
dependence of the background metric on our brane's location in the
fifth ($x_4$) dimension, or rather its distance, $r_c$, from the
other, and equally ``end--of--the--world'', 3-brane (also termed the
UV-brane). To be specific, one has, for the Higgs vacuum expectation
value (and, similarly, for the mass), $v = \tilde v \, \exp(- \pi \,
k_5 \, r_c)$, where $\tilde v = {\cal O}(M_5)$ and $k_5$ is a measure
of the bulk curvature.

With the extent of the hierarchy now being determined by the modulus
($r_c$) of the compactified fifth dimension, the latter must be
stabilized, an issue not addressed by the originators of the model. In
other words, if the modulus is construed to be a dynamical field
${\cal M}$, then a mechanism that forces the field to settle ( at $\langle
{\cal M} \rangle = r_{c}$) should exist and be operative. As
Ref.\cite{Goldberger:1999uk} showed, this could be achieved by
introducing a new scalar field $\phi$, with a non-vanishing potential,
in the five-dimensional bulk. As $\phi$ interacts with ${\cal
M}$ through the metric, integrating out the former would result in an
effective potential $V_{\rm eff}({\cal M})$.  An apt and simple choice 
of the scalar-potential alongwith boundary conditions (without any 
discernible hierarchy) can, then, lead to a 
suitable form for $V_{\rm eff}({\cal M})$ and, thereby, an 
 appropriate $r_c$. 

With gravity percolating into the bulk, it is obvious that
compactification would lead to a Kaluza-Klein (KK) tower of gravitons,
with masses given by $m_n = x_n \, k_5 \, \exp(- \pi \, k_5 \, r_c)$
where $x_n$ denote the roots of the Bessel function of order one. Both
the applicability of semi-classical arguments (upon which the model
hinges) as well as string theoretic arguments relating the D3 brane
tension to the string scale (and, hence, to $M_5$ through Yang-Mills
gauge couplings) restrict $k_5/M_5 \lapp
0.15$~\cite{Davoudiasl:1999jd}.  Thus, one expects the first KK-mode of
the graviton, to be, at best, a few times heavier than the Higgs
boson.  Furthermore, the very warping that explains the hierarchy also
concentrates the KK-modes (though, not the lowest and massless mode)
near the TeV brane, thereby enhancing their couplings to the SM
fields. Consequently, several search strategies at the LHC were
designed~\cite{ATLAS:2011ab,Aad:2012cy,ATLAS:2013jma,Khachatryan:2014gha}
to detect their signatures in a multitude of channels. Negative
results from the same~\ viz. $m_1 \gapp 2.66\tev$ (at 95\%
C.L.)~\cite{Aad:2015mna, EXO-12-045}, thus, impose severe constraints
on the model. It can be argued that, unless we allow a little
hierarchy in the ratio $\tilde v / M_5$, the RS scenario can be ruled
out as a solution to the SM hierarchy problem.

The situation improves significantly if one were to consider an
extension of the scenario to two extra dimensions with nested
warpings~\cite{Choudhury:2006nj}. The graviton spectrum, while now
being enlarged to a tower of towers, is different from that in the
five-dimensional case in two crucial aspects. For one, the change
wrought in the graviton wave function results in the mass of the first
KK-mode being significantly higher than that of the corresponding mode
in the (five-dimensional) RS case\footnote{This is easy to understand
  once one realizes that the resolution of the hierarchy between the
  electroweak scale and the fundamental scale is now shared between
  two warpings.  Consequently, the extent of the individual warpings
  is smaller here than required in the RS case.}.  As this happens for
natural values of the parameters, and does not need any fine-tuning,
this feature, on its own, would imply a weakening of the
aforementioned ``little hierarchy'' that the original RS scenario
needed so as to explain the nonobservation of gravitons at the
LHC~\cite{Arun:2014dga}.  More importantly, the large coupling (to the
SM fields) enhancement that allowed for the graviton KK-modes to be
extensively produced at the LHC, is now tempered to a great
degree~\cite{Arun:2014dga}, a consequence, once again, of the double
warping. Consequently, the graviton production rates are further
suppressed and the scenario easily survives the current bounds from
the LHC~\cite{Arun:2014dga,Arun:2015ubr}. On the other hand, 
while the allowed parameter space of the model is still quite
extensive, it can be probed well in the current run of LHC.

This, along with the fact that formulating the theory in a
six-dimensional world has many other benefits, especially when the SM
fields are also allowed into the
bulk~\cite{Arun:2015kva,Arun:2016ela}, renders this construction
rather interesting. In particular, with the 
  four-dimensional theory getting supplanted by a five-dimensional one
  at the lower of the two compactification scales, the infrared is 
  effectively screened from modes traversing the far ultraviolet. This 
  is also reflected by the explicit computations of the electroweak 
  precision variables\cite{Arun:2016ela}, which demonstrated that the 
  little hierarchy is no longer a major issue.
However, the very issue
of stabilizing the moduli (two in the current case, as opposed to a
single one in the 5-dimensional one) has not been addressed so
far. This assumes particular significance in that the structure
formulated in Ref.\cite{Choudhury:2006nj} does not boast of a
conformally flat geometry. Furthermore, the branes are not necessarily
flat and this introduces its own set of complications. In this paper,
we aim to rectify this situation and develop two related, but
distinct, stabilization mechanisms, somewhat analogous to those in
Refs.\cite{Goldberger:1999uk,Cline:2000xn,DeWolfe:1999cp,back_reaction}.
This would also be seen to offer a stabilization mechanism for the
modulus in a UED theory, thereby addressing a long-standing general
lacuna in this otherwise attractive scenario.

Before we venture into the actual stabilization mechanism or even a
detailed discussion of the scenario, we wish to clarify certain issues.
Naively, it might be argued that having the gravitons to be heavier
than in the RS would result in a worse fine tuning for the electroweak
scale.  As we have already mentioned, this moderate heaviness is but a
consequence of there being two extra dimensions. To appreciate this,
let us consider a sequence of unrelated scenarios. The first example
would be an ADD~\cite{ArkaniHamed:1998rs}-like scenario with two
extra-dimensions being compactified toroidally, with radii (possibly
different) only somewhat larger than $M_6^{-1}$, namely $R_i =
\theta_i \, M_6^{-1}$ with $\theta_i \gapp 1$.  This would have meant
$M_{\rm Pl}^2 = M_6^2 \, \theta_1 \, \theta_2$. In the analogous
five-dimensional theory, one would have, instead, $M_{\rm Pl}^2 =
M_5^2 \, \theta_1$. Thus, for the theory with the larger number of
extra dimensions, one would have a smaller hierarchy between the
fundamental $M_5$, $M_6$ etc. as the case may be and the electroweak scale. Consider, as the next example, a
simplistic generalization of the RS scenario to a slice of AdS$_6$
bounded by two 4-branes, such that the apparent scale on the IR-brane
is a few TeVs. This particular (multi-TeV) scale would, then, be
protected with the graviton KK-modes now lying at at the same
scale. If the 5-dimensional world be further compactified (or, even,
orbifolded) over a circle of small radius, there would extend an
additional factor in the relation between $M_6$ and $M_{\rm Pl}$,
thereby further ameliorating the hierarchy problem. As we shall see,
much the same happens in the present case.

On a related note, the cutoff of the effective 
four-dimensional theory needs to be identified too. For an 
effective theory, this is often described as the scale at
which the loop contributions (often very large) are
to be cut off, for the new physics beyond this scale would naturally 
regulate such contributions. Nonetheless, with the ultraviolet 
completion of the present theory being unknown (in the absence 
of any quantum theory of gravity), this cancellation cannot 
be demonstrated exactly. 
However, within the five-dimensional context, it has been 
argued that the addition of the Planck-brane
and/or the TeV-brane allows for a holographic interpretation, with 
the former acting as a regulator leading to a UV cutoff, of the order 
of the inverse of the modulus, on the corresponding conformal 
field theory~\cite{ArkaniHamed:2000ds,Rattazzi:2000hs,PerezVictoria:2001pa}.
A similar conclusion also holds for theories with gauge fields
extended in to the warped bulk~\cite{Davoudiasl:1999jd,Pomarol:1999ad,Agashe:2002jx}. While no such duality has been explicitly 
constructed for the six-dimensional
case, one such would obviously exist, for, in a certain limit, 
the bulk is indeed AdS$_6$--like. Thus, the branes would  
provide a regulator, albeit in a deformed CFT. 
In particular,  the cutoff for the four-dimensional quantum field 
theory is set not by $M_6$, but the inverse of the larger of the 
two moduli. At such a scale, the higher-dimensional nature of the theory 
becomes quite apparent, and the four-dimensional effective theory 
(including the graviton KK-modes) is no longer an apt language.
And while the compactification mechanism is not specified here 
(or within the RS theory), the physics
responsible for it must be incorporated in any description that
reaches beyond this scale.

\section{The 6D warped model} 
\label{6warped}
The space-time of interest is a six-dimensional one with 
successive (nested) warpings along the two 
compactified dimensions. The uncompactified directions support 
four-dimensional ($x^\mu)$ Lorentz symmetry while the compactified 
directions are individually $Z_2$-orbifolded. In other words, we have 
$M^{1,5}\rightarrow[M^{1,3}\times S^1/Z_2]\times S^1/Z_2$. 
Representing the compact directions by the angular
coordinates $x_{4,5} \in [0,\pi]$ with $R_y$ and $r_z$ being the
corresponding moduli, the 
line element is, thus, given by~\cite{Choudhury:2006nj}
\begin{equation}
\label{metric}
ds^2_6= b^2(x_5)[a^2(x_4)\eta_{\mu\nu}dx^{\mu}dx^{\nu}+R_y^2dx_4^2]+r_z^2dx_5^2 \ ,
\end{equation} 
where $\eta_{\mu \nu}$ is the flat metric on the 
four-dimensional slice of spacetime.
As in the RS case, orbifolding, in the presence of nontrivial warp
factors, necessitates the presence of localized energy densities at
the orbifold fixed points, and in the present case, these appear in
the form of tensions associated with the four end-of-the-world
4-branes.

Denoting the natural (quantum gravity) scale in six dimensions by $M_6$ and 
the negative (six dimensional) bulk cosmological constant by 
$\Lambda_6$, the total bulk-brane action is, thus,
\begin{equation}
\barr{rcl}
{\cal{S}}&=& \dis {\cal{S}}_6+{\cal{S}}_5 \\[1.5ex]
{\cal{S}}_6&=& \dis \int d^4x \, dx_4 \, dx_5 \sqrt{-g_6} \, 
   (M_{6}^4R_6-\Lambda_6)\\[1.5ex]
{\cal{S}}_5&=& \dis \int d^4x \, dx_4 \, dx_5 \sqrt{-g_5}\, 
      [V_1(x_5) \, \delta(x_4)+V_2(x_5) \, \delta(x_4-\pi)]\\
&+& \dis \int d^4x \, dx_4 \, dx_5\sqrt{-\tilde g_5} \, 
     [V_3(x_4) \, \delta(x_5)+V_4(x_4) \, \delta(x_5-\pi)] \ . 
\earr
\label{eq:6drs}
\end{equation}
The five-dimensional metrics in ${\cal S}_5$ are those induced on the
appropriate 4-branes which accord a rectangular box shape to the
space.  Furthermore, the SM (and other) fields may be localized on
additional 3-branes located at the four corners of the box, viz.
\[
{\cal{S}}_4 = \sum_{y_i, z_i = 0, \pi} \, 
\int d^4x \, dx_4 \, dx_5 \, 
\sqrt{-g_{4}} \, {\cal L}_i \, \delta(x_4- y_i) \, \delta(x_5- z_i) \ .
\]
Since ${\cal{S}}_4$ is not relevant to the discussions of this
paper, we shall not discuss it any further.

Rather than limit ourselves to the solutions to the Einstein equations
presented in Ref.\cite{Choudhury:2006nj}, we consider, here, a more
general class.  To motivate it, let us recollect that, in such models, the 
presence of bent branes is due to a ``lower-dimensional cosmological
constant'' induced on the brane. For example, the  four dimensional 
components of the Einstein equations, in the presence of such a 
term $\Omega$ would read
\[
a^2 \left[\frac{3}{R_y^2}\left(\frac{a''}{a}+\frac{a'^2}{a^2}\right)
             +\frac{2}{r_z^2}\left( 3 \dot b^2 + 2 b \ddot b
	           +\frac{\Lambda_6 \, r_z^2}{2 M_6^4}b^2\right)\right] 
             =\frac{\Omega}{r_z^2} \ ,
\]
where primes(dots) denote derivatives with respect to $x_4$ ($x_5$).
Introducing a constant of separation $ \widetilde{\Omega} $, 
we have 
\beq
\label{bequation}
3\dot b^2 + 2 b \ddot b +\frac{\Lambda_6 r_z^2}{2 M_6^4}b^2 = \widetilde{\Omega}
\eeq
and
\beq
\label{aequation}
\frac{3}{R_y^2}\left(\frac{a''}{a}+\frac{a'^2}{a^2}\right)
     +\frac{2}{r_z^2}\widetilde{\Omega} = \frac{\Omega}{r_z^2 \, a^2}\, .
\eeq
The first equation has the solution 
\beq
\label{generic_solz}
b(x_5) = b_1 \cosh(k |x_5| + b_2) \ , \qquad b_1 =
\sqrt{\frac{-\widetilde \Omega}{3 \, k^2}} = {\rm sech}(k \pi + b_2)
\ , \qquad k = r_z \sqrt{\frac{-\Lambda_6}{10 \, M_6^4}} \equiv r_z \,
M_6 \, \epsilon 
\eeq
 assuming\footnote{For $\widetilde \Omega > 0$,
  one would, instead, have $b(x_5) = \sqrt{\widetilde \Omega / 2 \,
    k^2} \, \sinh(k |x_5| + b_2)$.  Not much would change materially,
  except for the fact that $b_2 = 0$ would no longer be allowed unless
  one is willing to admit a vanishing metric, albeit only at a given
  slice of space-time. It is intriguing to note that the notion of a
  degenerate spacetime has received recent attention from a different
  standpoint~\cite{Kaul:2016lhx}.}  $\widetilde \Omega < 0$.  While
Ref.\cite{Choudhury:2006nj} had considered only the special case of
$b_2 = 0$, we shall admit the more general solution. As we shall see
below, a nonzero $b_2$ would have very important
consequences. Physically, $\widetilde \Omega$ (or equivalently $b_2$)
is related to the induced
cosmological constant on a five dimensional
hypersurface\footnote{It should be realized that a
    five-dimensional cosmological constant is very different from a
    four-dimensional one. Indeed, even for a large value of the
    former, one could be left with a vanishing value for the latter,
    as would be the case here.} along the constant $x_5$
direction. Differing values of $\widetilde \Omega$,
  thus, correspond to inequivalent extent of bending of the
  four-brane, and, hence, lead to different physical outcomes. We will
  demonstrate this shortly using widely different (in essence,
  limiting) values for $\widetilde \Omega$.  However, while the
  quantitative results do differ, qualitatively they turn out to be
  quite similar, with certain aspects essentially not changing at
  all. This was to be expected as many of the measurables (and
  certainly the most important ones) are only slowly varying functions
  of $\widetilde \Omega$. Consequently, the physical consequences
  (and the exact stabilization potential) of any arbitrary
  intermediate value of $\widetilde \Omega$ can be trivially obtained
  by effecting a simple interpolation between the results for the
  extremal values.

For future convenience, we 
have also introduced
  the dimensionless combination $\epsilon$. Clearly, for a
  semi-classical approach to be valid, the curvature must be
  significantly smaller than the mass scale of the theory. In other
  words, $\epsilon$ must be small\footnote{While a slightly larger 
  $\epsilon$ can be admitted, say by arguments relating the brane
tension to the scale of some underlying string theory
 (or even to $M_6$)~\cite{Davoudiasl:1999jd}, 
the applicability of the semiclassical approximation grows 
progressively worse. On the other hand, $\epsilon \lapp 0.15$
automatically ensures that the curvature in the $x_4$ -direction
is sufficiently small.}, namely $\epsilon \lapp 0.15$.
On the other hand, as we shall see below, too
  small an $\epsilon$ would either invalidate the resolution of the
  hierarchy problem, or, in the process, introduce a new (but smaller)
  hierarchy.

The solution to eqn.(\ref{aequation}) for a nonzero $\Omega$ is given
in terms of hyperbolic functions. While it is possible to work with
the general solution, the consequent algebra is exceedingly
complicated and the exercise does not proffer any extra insight that a
simplifying choice does not. As a nonzero $\Omega$ results in a
nonzero cosmological constant in the four-dimensional world, and as
the observed cosmological constant in our world is infinitesimally
small, we disregard it altogether and consider only\footnote{While
  this may be perceived as a fine-tuning, it is, at worst, exactly the
  same as that in the RS model. Indeed, $\Omega = 0$ is not a special
  solution, and the same argument could be made against any finite
  value for $\Omega$. On the other hand, $\Omega = 0$ could, in
  principle, be the result of some as yet unspecified symmetry
  \cite{Alexander:2001ic}.}  $\Omega = 0$. We do not claim to offer
any rationale for this choice but for the fact that it simplifies the
algebra for the rest of the article without losing any of the
essence. In this limit, the solution can be expressed as\footnote{Once
  again, we omit the second solution, viz. $e^{c x_4}$ for reasons
  analogous to those operative for $b(x_5)$.}
\beq
\label{generic_soly}
a(x_4) = a_1 \, e^{-c|x_4|}  \qquad   
    c \equiv b_1 \, k \, \frac{R_y}{r_z} \ .
\eeq
Normalizing the warp factors, at their maximum values, through
$a(0)=1$ and $b(\pi)= 1$, and imposing the orbifolding conditions, we
have
\beq
\label{sol_a}
b(x_5) = \frac{\cosh(k |x_5| + b_2)}{\cosh(k \pi + b_2)} \ ,
\qquad
a(x_4) = \exp(- c \, |x_4|) \ .
\eeq
The brane potentials are determined by the junction conditions. 
The ones at $x_5 = 0, \pi$ are simple  and 
are given by
\beq
\label{brane_potential_5}
V_3 = \frac{-8 M_6^4 k}{r_z} \tanh(b_2) \ , \qquad 
V_4 = \frac{8 M_6^4 k}{r_z} \tanh(k \pi + b_2) \ ,
\eeq
whereas the ones at $x_4 = 0, \pi$ have $x_5$--dependent tensions 
\beq
\label{brane_potential}
V_1(x_5) = - V_2(x_5) = 
 \frac{8 M_6^4 c}{R_y \, b(x_5)} 
   = \frac{8 M_6^4 k}{r_z} \, {\rm sech}(k |x_5| + b_2)\ .
\eeq

It should be noted that the Israel junction condition $V_1 = - V_2$ is
necessitated only by our focus on $\Omega = 0$, or, in other
words, a configuration wherein the four-dimensional cosmological
constant vanishes exactly. Had we admitted $\Omega \neq 0$, this
equality of magnitude would neither have been necessary nor would it
have held. This, of course, is exactly as in the RS case. The
dependence of $V_{1,2}$ on $x_5$ is easy to understand.  Each slice of
$x_5$ could, potentially, host a 4-brane, with distinct $(3 +
1)$-dimensional worlds at the ends.  Only if the potentials localized
at the end of the branes are equal and opposite and related to the
``overall size'' of the 5-dimensional metric in that slice (just as in
the RS case), would these hypothetical $(3 + 1)$-dimensional worlds be
associated with a vanishing cosmological constant.  As has been
demonstrated in Refs.\cite{Choudhury:2006nj, Arun:2016ela}, such a
$x_5$--dependent potential could be occasioned by a brane-localized
scalar field, such as a kink solution corresponding to a quartic
potential, or in a theory with a non-trivial kinetic term.

It should be noted, though, that with these particular forms for
$V_{1,2}$ are {\em not strict requirements} for the model. Such a
choice only helps to reduce the algebra. Indeed, as long as
eqn.(\ref{brane_potential}) holds at $x_5 = 0$ (with no restrictions
for $x_5 \neq 0$), the vanishing of the four-dimensional cosmological
constant is guaranteed.  However, the relaxation of
eqn.(\ref{brane_potential}) does not add anything qualitatively
different to either the phenomenology (whether in the graviton
sector~\cite{Arun:2014dga,Arun:2015ubr} or in the SM
sector\cite{Arun:2015kva,Arun:2016ela}) or to the main thrust of this
paper, namely the stability of the scenario.

We now turn to the consequences of choosing a particular value for 
$b_2$ (this choice, as we shall see later, also serves to determine 
$c$). Rather than discuss the generic case (which does not afford 
closed-form analytical solutions), we illustrate the situations for 
two extreme limits. Physically, one of the limits corresponds to a 
vanishing five-dimensional cosmological constant (equivalently, 
straight, or unbent, four-branes at the ends of the world). The 
opposite limit corresponds to the case wherein
the four-branes suffer the maximum possible bending commensurate with a 
semiclassical analysis (or, in other words, a five-dimensional cosmological 
constant comparable to the fundamental scale). 
The low energy phenomenology, naturally, would 
turn out to be quite different in the two cases. Clearly, any intermediate 
value of $b_2$ would correspond to an intermediate value of the five-dimensional
cosmological constant and, similarly, for the low-energy phenomenology.

\noindent
\underline{\bf Case 1:} The situation of $b_2 = 0$ recovers the results of
Ref.\cite{Choudhury:2006nj} and we have 
\beq \barr{rcl} c & = & \dis
\frac{R_y \, k}{r_z} \, {\rm sech}(k \pi) \\[2ex] 
V_1(x_5) &= &\dis -
V_2(x_5) = \frac{8 M_6^4 k}{r_z} {\rm sech}(k |x_5|) \, ,
\\[2ex]
V_3 &= & 0 \\[1ex] V_4 & = & \dis \frac{8 M_6^4 k }{r_z} \, \tanh(k \pi) \, \, .
\earr \eeq 
This, obviously, corresponds to a bent brane scenario with
  nonvanishing induced five-dimensional cosmological constants on 
  the hypersurfaces at $x_5 = 0,\pi$. This could easily be seen 
  by observing that the induced
  metric on the $x_5 = 0$ surface, apart from an overall $b(0)$ factor, is given by
\[
ds_5^2 = e^{- 2 c |x_4|} \eta_{\mu \nu} dx^{\mu}dx^{\nu} + R_y^2 dx_4^2 \, \, ,
\]
or, in other words,
  the induced geometry is $AdS_5$-like.


\noindent
\underline{\bf Case 2:} 
In the opposite limit, viz. $b_2 \to \infty$, we have $b(x_5)
  \approx (b_1 / 2) \exp(k |x_5| + b_2)$ and, hence, the normalization
  of the warp factor would imply $b_1 \approx 2 \, \exp(- k \pi - b_2)
  \to 0$.  Consequently, one is forced to $c \ll k$, unless one were
  to admit a large, and unpleasant, hierarchy between $R_y$ and
  $r_z$. This situation should be contrasted to the previous case,
  where the limit was realizable for both branches of the theory,
  viz. $c \ll k$ as well as a moderate $c > k$.

With $c \rightarrow 0$ the brane potentials now read
\beq 
\barr{rcl}
V_1 = - V_2 & \approx & 0 
\\[1ex] 
V_3 & \approx & \dis \frac{- 8 M_6^4 k}{r_z}  
\approx - V_4 
\earr
\eeq 
The fact of $V_3 \approx - V_4$ reveals the near vanishing of brane-induced
cosmological constant. As for the 
line element,  in this limit, 
\[
ds^2 = e^{2 k (|x_5|-\pi)} \Big( e^{- 2 c |x_4|} \eta_{\mu \nu} dx^{\mu}dx^{\nu} +R_y^2 dx_4^2 \Big) + r_z^2 dx_5^2 
\]
\[
\approx e^{2 k (|x_5|-\pi)} \Big(\eta_{\mu \nu} dx^{\mu}dx^{\nu} +R_y^2 dx_4^2 \Big) + r_z^2 dx_5^2 
\]
or that the metric is nearly conformally flat. It should be 
realized, though, that the approximate conformal flatness would have 
followed as long as $c \ll k$ (i.e., for $k \gapp 10$) and did not need 
$b_2 \to \infty$. However, a finite value of $b_2$ would have translated 
to unequal brane tensions and, consequently, nonvanishing induced 
cosmological constants.

The two opposing limits of $b_2$ are not special, but only 
  serve to simplify the algebra. Any intermediate value of $b_2$ would 
  only lead to phenomenological situations that interpolate between 
  those listed above. In the following, we shall detail not only the 
  stabilization of the radii, but also that of $b_2$.

\section{Radii Stabilization}

While it may seem, at first sight, that moduli stabilization in this
(6D) framework can proceed in a fashion identical to that in the RS
paradigm, there are certain crucial differences. In particular (and,
as we shall see below), if we attempt a naive GW
\cite{Goldberger:1999uk}--like mechanism, only one combination of
the two moduli can be stabilized. This is but a reflection of the
well-known fact that, for a multidimensional hidden compact space,
it is easier to stabilize the shape rather than the volume.  It
should be realized, though, that had we been interested in a
different compactification (such as, for example, $M^{(1,3)} \times
S^2$ with an appropriate orbifolding), a single-field GW-like
mechanism would indeed be enough. This is as expected, for in such a
case there would, but, be only one modulus to stabilize. However,
such a compactification is not favoured phenomenologically as, on
the one hand, it requires extra fields to counterbalance the
curvature of $S^2$, while, on the other, if the SM fields are
extended in to the bulk (so as to fully exploit the advantages of
the 6D construction), the resultant spectrum cannot, easily, be made
consonant with low energy observations.

While the same mechanism would work irrespective of the choice for
induced cosmological constant, the algebraic simplification is
significant in the two limits discussed in the preceding section.
Similarly, treating the two distinct regimes (viz small $k$ and large
$k$) separately brings forth an appreciation of both the overall
mechanism, as well as the subtle differences in the implementation
thereof.

Before we do this, though, let us reexamine some potentially confusing
features of this scenario, in particular the roles of the brane
localized potentials $V_i$, the separation constant $\widetilde
\Omega$ and the constant $b_2$. At first glance, the ``choices'' might
seem to associated with fine-tunings. We begin by showing that not all
of them are independent and, then, explore the stabilization of the
truly independent.

To begin with, it should be realized that the special 
case of $\widetilde \Omega = 0$ would have led to a generic solution 
of the form 
\[
b(x_5) = \beta_1 \, \cosh^{2/5}\left[ \frac{5 k x_5}{2} + \beta_2 \right]
\]
where $\beta_{1,2}$ are the constants of integration, with
  $\beta_1$ to be fixed by our normalization of $b(x_5 = \pi) = 1$.
This special solution is unique to $\widetilde \Omega = 0$ and
untenable for $\widetilde \Omega \neq 0$, when only the solution of
eqn.(\ref{generic_solz}) applies. More importantly, the two solutions differ by at most
  $50\%$ (almost independent of the value of $\widetilde \Omega$). 
  For large $k$ ($\sim 8$, as would be the case for preferred 
    solution for the hierarchy problem), the warp factors are very
  nearly indistinguishable, throughout the bulk, with the 
    difference being noticeable only very close to the IR brane.
In other
words, the conclusions that we would draw are not very sensitive to
the exact value of $\widetilde \Omega$. Put differently, there is no
severe fine-tuning associated with $\widetilde \Omega$.

Note further that  eqn.(\ref{generic_solz}) also implies 
\[
   \widetilde \Omega = - 3 \, b_1^2 \, k^2 
   = - 3 \, k^2 \, {\rm sech}^2(k \, \pi + b_2)
\]
whereas eqn.(\ref{brane_potential_5}) 
\[ 
   V_3  =  -8 \, \sqrt{\frac{-\Lambda_6 \, M_6^4}{10 }} \, \tanh b_2 \ , 
   \qquad \quad 
   \frac{- V_4}{V_3} = \frac{\tanh(k \pi + b_2)}{\tanh b_2}\ .
\]
In other words, there is a one-to-one relation between $(V_3, V_4)$
and $(k, b_2)$ or, equivalently, $(k, \widetilde \Omega)$. Stabilizing
one set automatically stabilizes the others. While we propose below 
a mechanism to stabilize the last (or, equivalently, the first) set, 
note that we have already seen that the dependence of physical observables 
on $\widetilde \Omega$ is a suppressed one. Thus, stabilizing $k$ would 
be enough. 

\subsection{Small $k$ and large $c$}
As we have discussed in Sec.\ref{6warped}, in this regime, the metric
cannot be approximated by a conformally flat one, and, of the two
limits discussed therein, only Case I can be applicable. Rather than
work with the general solution, we shall work in this limit, for it
simplifies the algebra considerably without altering the physical
essence.

As we have also explained earlier, starting with a single canonically
quantized scalar field, it is not possible to stabilize both the
moduli. Consequently, we postulate two such scalar fields.  In order
to minimize the number of effective four-dimensional fields (on KK
reduction), we incorporate one scalar field $\phi_1 (x_\mu, x_4,x_5)$,
permeating the entire bulk, that would serve to stabilize $r_z$ (or,
equivalently, the dimensionless quantity $k$). A second field
$\phi_2(x_\mu, x_4)$, introduced (localized) only on the $x_5=0$
brane, would, similarly, stabilize the length ($R_y$) of the
brane. Given the box structure and the orbifolding, together, they
stabilize both the moduli.
 
The Lagrangians for these scalars are given by
\beq
\label{z-scalar}
{\cal L}_6 = \sqrt{-g} \Big(-\frac{1}{2} g^{MN}\partial_M \phi_1 \partial_N \phi_1 -\frac{1}{2}m^2 \phi_1^2 \Big) + \sqrt{-g_5} \Big(U_1(\phi_1)\delta(x_5) +U_2(\phi_1)\delta(x_5-\pi)\Big)
\eeq
and
\beq
\label{y-scalar}
{\cal L}_5 = \sqrt{-g_5} \Big(-\frac{1}{2} g^{\bar{M}
  \bar{N}}\partial_{\bar{M}} \phi_2 \partial_{\bar{N}} \phi_2
-\frac{1}{2}m^2 \phi_2^2 \Big)\delta(x_5) + \sqrt{-g_4}
\Big(U_3(\phi_2)\delta(x_4) +U_4(\phi_2)\delta(x_4-\pi)\Big)\delta(x_5) 
\eeq
respectively. Here, $g = {\rm det}(g_{MN}) = -a^4 b^5 R_y r_z$,
whereas, for the induced metrics, we have
$g_5 =
{\rm det}(g_{\bar{M}\bar{N}})= - R_y a^4b^5 $ and $g_4 = {\rm det}(g_{\mu\nu}) 
= - a^4 b^4$.
 
In particular, the 
5d metric, apart from the constant $b^2(0)$, induced on the $x_5 = 0$ brane is 
\[
ds_5^2 = e^{-2 c |x_4|} \eta_{\mu \nu}dx^{\mu}dx^{\nu} + R_y^2 dx_4^2 \, .
\]
Given this $AdS_5 $ geometry and the form of ${\cal L}_5$, it is clear
that the stabilization of $R_y$ can
proceed exactly as in the GW
mechanism~\cite{Goldberger:1999uk} or its variants~\cite{Cline:2000xn}
using the classical configuration of $\phi_2$.  Since this technique
is well-known, we, for the sake of brevity, will eschew any details
here, assuming that $R_y$ can be stabilized.  Indeed, with an
appropriately modified five-dimensional potential for $\phi_2$, it is
also possible to take into account the back reaction and achieve an
exact solution\cite{DeWolfe:1999cp}. We will come back to a 
generalized version of this.

Unlike the 4-brane at $x_5 = 0$ (hereafter called the 
$4_0$ brane) itself, the $x_5-$direction possesses a
non-zero induced cosmological constant, as shown in section \ref{6warped}. 
Stabilization of this
direction, thus, requires a more careful 
analysis which we proceed to now.
 
The effective potential for $r_z$ (equivalently, $k$) can be obtained,
starting from the Lagrangian of eq. \ref{z-scalar}.  While the
classical configuration of $\phi$ could, in principle, have nontrivial
dependences on both $x_4$ and $x_5$ (and, yet, maintain the requisite
Lorentz symmetry), such a general consequence would have required
complicating the boundary-localized terms and does not add anything
qualitatively different to the system. Since we are primarily
interested
  in the effective potential in the $x_5$-direction, for brevity's
sake, we restrict our discussion to the case where $\phi$ has a nontrivial dependence only along $x_5$ and denoting it
\[
\langle \phi_1(x_\mu, x_4,x_5) \rangle = \frac{\phi(x_5)}{\sqrt{R_y r_z}} \ ,
\]
the effective one-dimensional Lagrangian for $\phi(x_5)$ is given by
\beq
\widehat{{\cal L}_6} = \frac{a^4b^5}{2}
    \left[-r_z^{-2} (\partial_5\phi)^2 -m^2\phi^2\right] 
     +  a^4 b^5 R_y\left[U_1(\phi_1)\delta(x_5) +U_2(\phi_1)\delta(x_5-\pi) \right] \, .
\label{one-dim_lag}
\eeq
Understandably, $a(x_4)$ appears only as an overall multiplicative 
factor and plays no dynamical role. The corresponding equation of motion is
\beq
\label{eqphi}
\partial_5(b^5\partial_5 \phi) - b^5 m^2 r_z^2 \phi + R_y r_z^2 b^5\Big(\frac{\partial U_1(\phi_1)}{\partial \phi}\delta(x_5)+\frac{\partial U_2(\phi_1)}{\partial \phi}\delta(x_5-\pi) \Big) = 0  \, .
\eeq
The solution, in the bulk, is given in terms
of associated Legendre functions, viz. 
\beq
\phi = {\rm sech}^{5/2}(k x_5) \, 
     \left[ c_1 P^{\nu}_{3/2}(\tanh(k x_5))+c_2 Q^{\nu}_{3/2}(\tanh(k x_5))\right]
 \, ,
\eeq
where $c_{1,2}$ are the constants of integration and 
\beq
\nu = \frac{5}{2}\sqrt{1 + \frac{4 \mu^2}{25}} \ , 
\qquad \mu \equiv  \frac{m r_z}{k} = \frac{m}{M_6 \, \epsilon}
\ .
\eeq
The constants $c_{1,2}$ can be determined once boundary conditions are
imposed.  To do this, we turn to the brane-localized potentials
$U_{1,2}(\phi_1)$ which, until now, were unspecified.  We are not
sensitive to the exact form of $U_{1,2}(\phi_1)$ as long as they admit
nonzero minima at $\phi = v_{1,2}$ respectively. Noting that the cutoff scale
on this brane is given by $R_y^{-1}$, such minima, for example, can be
easily achieved if one were to consider
$U_{1,2}(\phi_1) = V_{3,4} + R_y^{-1}\lambda_{1,2} \Big(\phi^2 -
v_{1,2}^2 \Big)^2$ with $\lambda_{1,2}$ being dimensionless constants
and $V_{3,4}$ being defined as in
eq.(\ref{eq:6drs}).\footnote{On the boundary, once the scalar field
  $\phi$ settles down to the vacuum $v_{1,2}$,
  the brane localized potential becomes $U_{1,2}=V_{3,4}$ and one
  recovers the action given in eq.(\ref{eq:6drs}). While this
    unifies the explanation of the brane-tensions $V_{3,4}$ with the
    stabilization mechanism, truthfully, it, of course, does not yet
    explain their values. On the other hand, as we have explained
    earlier, with Einstein's equations and matter equation of motion
    being coupled, only certain values of $V_{3,4}$ can be consistent
    with the metric and the orbifolding. The stabilization of $k$,
    though, would imply the stabilization of $V_{3,4}$ too.
  } This immediately
leads to
\[
\barr{rcl}
c_2 & = & \dis
\frac{v_1P^{\nu}_{3/2}(\tau_\pi)-v_2 \cosh^{5/2}(k \pi) P^{\nu}_{3/2}(0)}
     {Q^{\nu}_{3/2}(0)P^{\nu}_{3/2}(\tau_\pi)-Q^{\nu}_{3/2}(\tau_\pi)P^{\nu}_{3/2}(0)}
\\[2ex]
c_1 & = & \dis \frac{1}{P^{\nu}_{3/2}(0)}\Big(v_1 - c_2 Q^{\nu}_{3/2}(0)\Big)
\\[2ex]
\tau_\pi & \equiv & \tanh(k \pi) \ .
\earr
\]
Putting the solution back in the effective
Lagrangian~\ref{one-dim_lag}, we have
\[
\barr{rcl}
\widehat{\cal L}_6 & = & \dis \frac{k^2 a^4}{2 \, r_z^2} \, 
{\rm sech}^5(k \pi) \,
     \left[ \frac{(5-2 \nu)^2}{4} 
  \Big(c_1 P^{\nu}_{5/2}(\tanh(k x_5))+c_2 Q^{\nu}_{5/2}(\tanh(k x_5))\Big)^2  
  \right.
\\[1ex]
& & \dis \hspace*{8em}
\left. -\mu^2 \Big(c_1 P^{\nu}_{3/2}(\tanh(k x_5))+c_2 Q^{\nu}_{3/2}(\tanh(k x_5))\Big)^2 \right] \ .
\earr
\]
Eliminating the irrelevant factor $a^4(x_4)$ and integrating $\widehat
{\cal L}_6$ over $x_5$, we would obtain an effective potential for
$k$, defined, in dimensionless form, as
\beq
  V_{\rm eff}(k) \equiv \frac{1}{M_6^2 \, v_1^2} \, 
                    \int dx_5 \frac{\widehat{\cal L}_6}{a^4(x_4)} \ .
\eeq
Since $k = r_z \, M_6 \, \epsilon$, with the last two quantities being
fixed parameters of the theory, $V_{\rm eff}$ is, thus, equivalently,
a potential for $r_z$. As a closed form expression for $V_{\rm eff}$
is not possible, and even a good approximate form complicated enough,
we present it, instead, only in a graphical form.
\begin{figure}[!h]
\vspace*{-20pt}
\epsfxsize=7cm\epsfysize=10cm 
\epsfbox{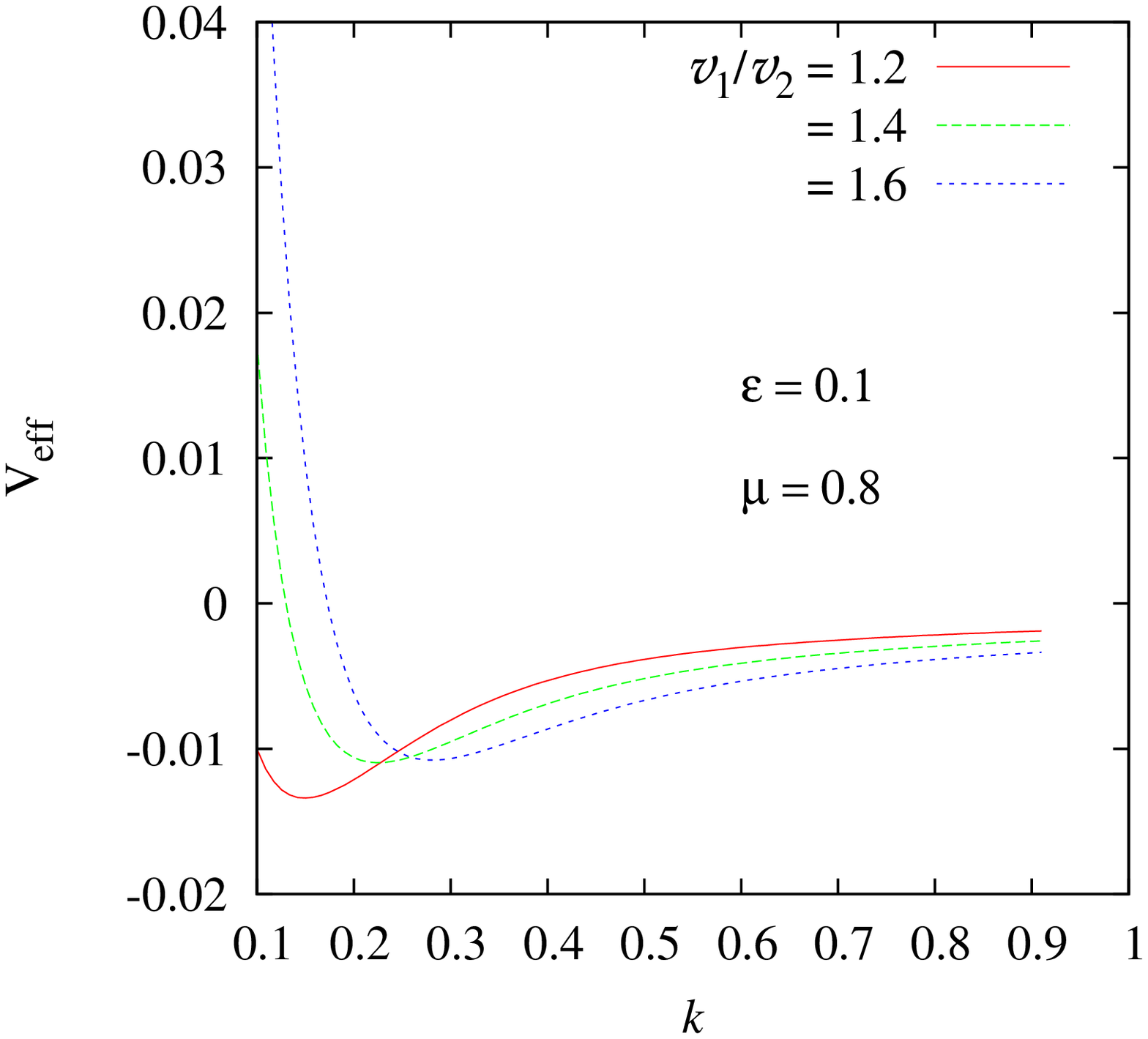}
\epsfxsize=7cm\epsfysize=10cm 
\epsfbox{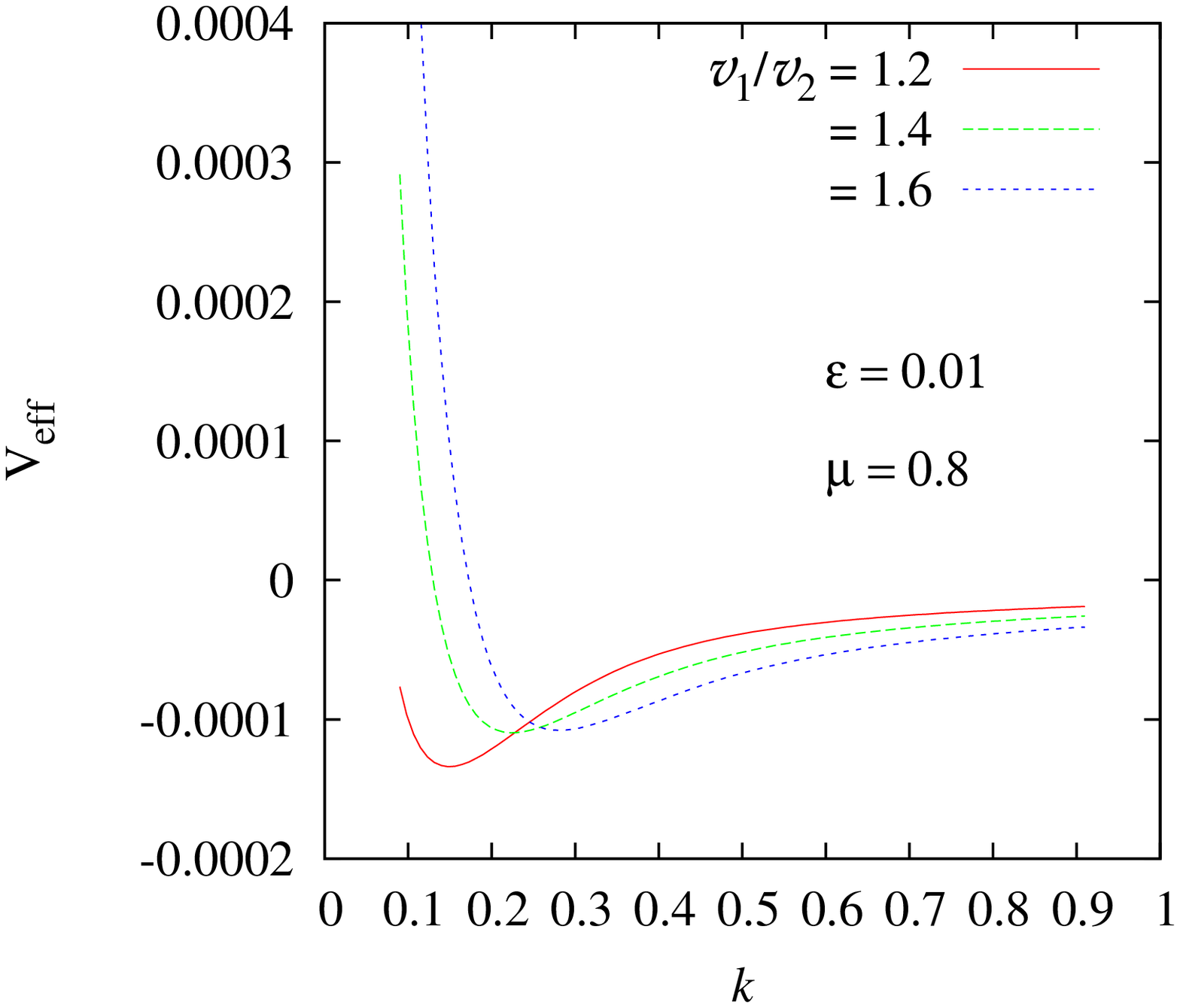}
\vspace*{-50pt}
\caption{\em 
The effective potential $V_{eff}(k)$ for different values of the 
ratio $v_2/v_1$ of the classical values of the field $\phi$ on the 
two constant-$x_5$ branes. The left (right) panels correspond to 
$\epsilon = 0.1 \, (0.01)$.}
\label{fig_1}
\end{figure}

In Fig.\ref{fig_1}, we display $V_{\rm eff}(k)$ for a fixed value of the 
mass parameter $\mu$ (equivalently, $m$). As is obvious, depending on
the ratio $v_2/v_1$, minima exist for $0.1 \lapp k \lapp 0.6$, the
range that is of particular interest not only to explain the
non-observation (so far) of the KK-graviton at the
LHC~\cite{Arun:2014dga, Arun:2015ubr}, but also for scenarios wherein
the SM fields are extended in to the bulk~\cite{Arun:2015kva,
Arun:2016ela}. What is particularly encouraging is that {\em such
minima arise for very natural values of the parameters} and are not 
overly sensitive to their precise values. Indeed, the strongest dependence, 
of the stabilized value of the modulus $r_z$, is on the ratio $v_1/v_2$ of the 
classical values.

\begin{figure}[!h]
\vspace*{-20pt}
\centering
\epsfxsize=7cm\epsfysize=10cm 
\epsfbox{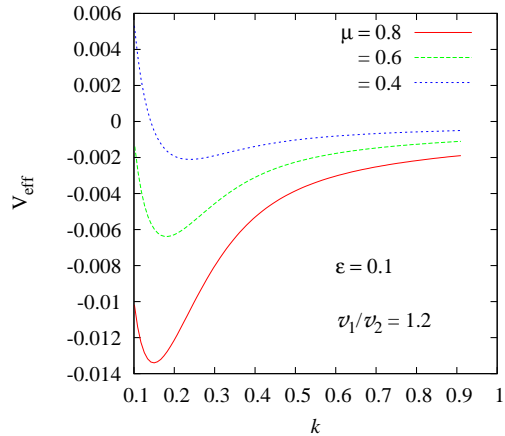}
\vspace*{-50pt}
\caption{\em The effective potential $V_{eff}(k)$ for different values of 
the mass of the bulk scalar $\phi_1$.}
\label{fig_mudep}
\end{figure}

It is also instructive to examine the dependence of  $V_{\rm eff}(k)$
on $\epsilon$ (as depicted in the two panels of Fig.\ref{fig_1}) and
$\mu$ (as shown in Fig.\ref{fig_mudep}). As can be readily
ascertained, while the size of the potential has a strong dependence
on $\mu$ (understandable, since it is $\mu$ that allows for a
nontrivial $V_{\rm eff}$), the position of the minimum has only a
muted dependence. 

Until now, we have neglected the back reaction on the metric 
due to the scalar field. While we could, in principle, attempt this, 
as we shall indeed do for the other regime (namely, large $k$) in 
the next section. However, in the current context, the presence of 
a non-negligible induced cosmological constant $\Lambda_{\rm ind}$ 
queers the pitch. In the absence of 
$\Lambda_{\rm ind}$, the
change in the warp factor due to back-reaction can be
computed exactly by integrating three 
first order equations (namely, the one for the scalar field, the warp factor
and the bulk potential).
For a non-zero $\Lambda_{\rm ind}$,
additional nonlinearities , that couple these 
three equations in a non-trivial manner, emerge~\cite{DeWolfe:1999cp}, 
and a closed-form analytic solution is not possible. Of course, the 
equations can still be solved numerically. However,
in view of the fact that such solutions can be easily obtained 
by deforming the solutions presented above, we eschew a detailed discussion for the sake of brevity.
The neglect of the back-reaction is eminently justified, for
the largest
value of the scalar field mass that we have used is sufficiently
smaller than $\Lambda_6^{1/6}$ (and, certainly $M_6$).  Consequently,
 the energy
content in the $\phi$-field is rather subdominant to that due to the bulk
cosmological constant and the back-reaction in the bulk is not much of a worry. 
Furthermore, with the hierarchy, for the most part,
being dictated by the warp factor in the $x_4-$direction, 
even moderate changes in $b(x_5)$, as would be introduced 
by the back-reaction, have relatively little bearing on the 
phenomenology.

\subsection{Large $k$ and small $c$}
In this regime, 
the metric is nearly conformally flat. With $c$
being infinitesimally small, neglecting the $c-$dependence of the metric
would not introduce a decipherable difference in the analysis. To
simplify the algebra, we will take recourse\footnote{Once again, the
choice of Case 2 does not represent a fine-tuning. Rather, it only 
serves to simplify the algebra permitting an analytics closed-form solution.}
  to \underline{\bf Case 2}
of section \ref{6warped}, whence the metric reduces to
\beq
ds^2 = e^{2 A(x_{5})}\Big(\eta_{\mu \nu}dx^{\mu}dx^{\nu} + R_y^2 dx_4^2  \Big) + r_z^2 dx_5^2  \, ,
\label{line_elem_large_k}
\eeq
where in the absence of backreaction $A(x_5) = k |x_5|$.  
With the $4_0$-brane-localized induced cosmological constant being 
infinitesimally small, it is possible to obtain an almost exact
solution incorporating the back reaction as well and we now attempt
this. Introducing a scalar field $\phi$ in the bulk, the entire action is
given by\footnote{We do not write $\Lambda_6$ explicitly, preferring to 
include it in $V(\phi)$.}
\beq
\label{action}
S= \int d^6x \sqrt{-g}\left[
   M_6^4 \, R - \, \frac{1}{2}(\partial \phi)^2-V(\phi)\right] \, \, .
\eeq
The corresponding equations of motion are
\beq
\barr{rcl}
\label{eqofmotion}
\ddot \phi + 5 \dot \phi \, \dot A & = & \dis 
            r_z^2 \, \frac{\partial V}{\partial \phi}
\\[1ex]
5 \dot A + 2 \ddot A & = & \dis
\frac{-r_z^2}{2 \, M_6^4} \,
\left[ \frac{\dot \phi^2}{2 \, r_z^2} + V(\phi) \right]
\\[2ex]
{\dot A}^2 & = & \dis 
\frac{r_z^2}{10 \, M_6^4}\left[ \frac{\dot \phi^2}{2 \, r_z^2} \, - 
         V(\phi)\right] \ .
\earr
\eeq
For a scalar with a localized potential on the $x_5-$constant 4-branes
$V(\phi)$ could be written as
\[
V(\phi) = V_{\rm bulk}(\phi) + r_z^{-1} \, 
      \left[ f_{0}(\phi(0)) \, \delta(x_5) 
           + f_{\pi}(\phi(\pi)) \, \delta(x_5 - \pi) \right]
\]
where $V_{\rm bulk}(\phi)$ is the bulk potential and 
$f_{0,\pi}(\phi(x_5))$ are some as-yet undetermined
functions of the scalar field.
Integrating eqn.(\ref{eqofmotion}) across the 4-brane 
locations ($\alpha \equiv x_5 = 0,\pi$),  we have
\[
\barr{rcl}
\dis \dot A \, \Big|_{\alpha-\epsilon}^{\alpha+\epsilon} & = & \dis 
        \frac{-1}{4 M_6^4} \, f_\alpha(\phi(\alpha))
\\
\dis \dot \phi \, \Big|_{\alpha-\epsilon}^{\alpha+\epsilon} & = & \dis
r_z \frac{\partial f_\alpha}{\partial \phi}(\phi(\alpha)) \, \, \, ,
\earr
\]
which provide the junction conditions. An exact closed-form solution to
eqns.(\ref{eqofmotion}) can be obtained only for particular bulk potentials. 
Borrowing from techniques of supersymmetric quantum mechanics, we assume
the bulk potential can be expressed as 
\beq
\label{superpotential}
V_{\rm bulk} = \frac{1}{2}\Big( \frac{\partial W}{\partial \phi}  \Big)^2 - \frac{5}{2 M_6^4} W^2 \, \, \, ,
\eeq
where $W(\phi)$ can be thought of as a superpotential. This, immediately 
leads to
\beq
\barr{rcl}
\label{warpfactor}
\dot A & = & \dis \, \frac{r_z}{2 \, M_6^4}W
\\[1ex]
\dot \phi & = & \dis - \, 2 \,  r_z \frac{\partial W}{\partial \phi} 
\earr
\eeq
as long as $W(\phi)$ satisfies the junction conditions
\[
\barr{rcl}
\dis W \, \Big|_{\alpha-\epsilon}^{\alpha+\epsilon} 
     & = & \dis \frac{1}{2} \, \frac{1}{r_z} \, f_\alpha(\phi(\alpha))
\\[2ex]
\dis \frac{\partial W}{\partial \phi}\, 
     \Big|_{\alpha-\epsilon}^{\alpha+\epsilon} 
   & = & \dis -\, \frac{1}{2} \, \frac{\partial f_\alpha(\phi(\alpha)}{\partial \phi} \, \, \, .
\earr
\]
Each choice for $W(\phi)$ gives a different $V_{\rm bulk}$, but an 
analytic closed-form solution can be found for only some.  An
explicit example
is afforded by a quadratic superpotential~\cite{DeWolfe:1999cp,back_reaction}, namely
\[
W(\phi) = 2 \, M_6^5 \, \epsilon - \frac{1}{4} u \, M_6 \, 
            \phi^2 \ ,
\]
where $u \lapp 0.1$  is a constant, parameterizing 
not only the mass of $\phi$, but its (quartic) self-interaction as well. The 
corresponding brane localized potentials read
\[
\barr{rcl}
f_0(\phi) & = & \dis \, \frac{1}{2 \, r_z} \, W(\phi)  -\, \frac{1}{2} \, \frac{\partial W}{\partial \phi} \, (\phi - v_0 ) 
           + \gamma_0^2 \, (\phi - v_0)^2 
\\[1.5ex]
f_\pi(\phi) & = & \dis \, \frac{1}{2 \, r_z} \,  W(\phi)  -\, \frac{1}{2} \,\frac{\partial W}{\partial \phi} \, (\phi - v_\pi ) 
           + \gamma_\pi^2 \, (\phi - v_\pi)^2 \, \, \, ,
\earr
\]
where $\gamma_{\pi,0}$ are arbitrary positive constants that ensure that
$\phi(x_5)$ assumes values $v_{\pi,0}$ on the Planck (TeV) branes.
The solutions to eqns.(\ref{warpfactor}) are given by
\beq
\barr{rcl}
\phi(x_5) & = & \dis \phi_0 \, \exp \left(u \,M_6 \,r_z \, |x_5| \right)
\\[1ex]
A(x_5) & = & \dis k \, |x_5| - \frac{v_0^4}{8 \, M_6^4} \, 
            \exp \left(2 \, u \,M_6 \,r_z \, |x_5| \right) \, \, \, .
\earr
    \label{soln_large_k}
\eeq
Note that the warp factor has changed from the simple exponential form 
that it had in the absence of $\phi$.

It is worthwhile to reflect on the difference 
between this analysis and that presented in the preceding 
subsection. While we could have adopted the same 
procedure, namely substitute eqn.(\ref{soln_large_k}) in
eqn.(\ref{action}) and integrate over $x_5$ to yield an effective
potential $V_{\rm eff}(r_z)$, it is not necessary to do so. Rather,
note that the very structure of the solution (eqn.\ref{soln_large_k}),
along with the boundary-localized potential ensures that
\beq
  r_z = \frac{1}{u \, \pi \, M_6} \, \ln\frac{v_\pi^2}{v_0^2} \, \, \, .
\eeq
No other value for $r_z$ would admit a solution, consistent 
with the boundary conditions, to the system of 
coupled nonlinear differential equations that we are endowed with. It is 
also worthwhile to note that a natural set of values for $v_\pi / v_0$ and 
$u$ can reproduce the required $r_z$ (and, hence, the correct warping) without 
any fine-tuning being needed.

We now turn to the stabilization of $R_y$. If the approximation of
eqn.(\ref{line_elem_large_k}) were truly exact, $R_y$ cannot be
stabilized. On the other hand, it need not be, at least in the context
of hierarchy stabilization, for it really does not play a role in
defining the overall warp-factor. Apparently, thus, the primary
constraints would be those on the ADD
scenario~\cite{ArkaniHamed:1998rs}, such as deviation from Newton's
law or the fast cooling of a supernova. And while it might be argued
that such an extremely large value for $R_y$ reintroduces a hierarchy,
it is not obvious that this is a problem (far less a serious one),
given that $R_y$ plays only a subservient role in defining the gap
between $M_6$ and the electroweak scale. Indeed, well before $R_y$
becomes so large (the sub-millimeter range), $c$ becomes quite
non-negligible. This not only invalidates the approximation of
eqn.(\ref{line_elem_large_k}), but also carries the seed for the
stabilization of $R_y$.

The latter can proceed, for example, in a fashion exactly analogous to
the GW mechanism as defined for the original RS scenario. Consider,
for example, a second scalar $\phi_2$ (of mass $m_2 \lapp M_6$)
confined to the 4-brane at $x_5 = \pi$. Assume that the only
self-interactions are localized at the boundaries, viz. at $(x_4, x_5)
= (0, \pi)$ and $(\pi, \pi)$ which, in turn, force $\phi_2(0,\pi) =
v_3$ and $\phi_2(\pi,\pi) = v_4$. Clearly, this would lead to a
stabilized $R_y^{-1} \sim {\cal O}\left(m_2 \, \ln (v_3/v_4) \right)$,
and, consequently, to a moderate $R_y / r_z$ and a small $c$ (as desired). 

A more interesting option would be to locate $\phi_2$ on the $4_0$ brane
instead, with the boundaries now corresponding to $(x_4, x_5) = (0,
0)$ and $(\pi, 0)$ respectively. With $m_2$ now suffering a large
warping (due to $b(x_5)$), the stabilized value for $R_y^{-1}$ would,
naturally, be in the TeV range. This, immediately, raises the
intriguing possibility that new physics at a few-TeV scale could indeed
be stabilized by the SM Higgs itself (or a cousin of its). Even more
intriguingly, if one allows the SM fields to percolate into the $x_4$
direction, the setup under discussion would provide a dynamical
justification for the scale in a Universal Extra Dimension-like
scenario~\cite{Appelquist:2000nn}.

\section{Conclusion}
The six dimensional warped scenario provides a cure for various
ailments of Randall-Sundrum model. Nevertheless, the problem of
modulus stabilization, which was quite simple for Randall-Sundrum
scenario, had, until now, not been executed for either of the two
moduli in the nested warped model, largely on account of the fact that
the model's space-time is neither conformally flat, nor are the
end-of-the world branes flat. Consequently, the stabilization
mechanism presents a technical challenge, and this is the issue that
we have addressed in this paper.  To this end, we begin by exploring
the metric for nested warping, showing that the solutions for each of
the two regimes allowed to the theory can be generalized beyond what
was considered earlier.

In the small $k$ (equivalently, large $c$) regime of the theory, the
induced geometries on the $x_5$--constant 4-branes are AdS$_5$-like,
and hence a Goldberger-Wise mechanism (or even one incorporating
back-reaction) involving a brane-localized scalar field trivially
stabilizes the corresponding modulus $R_y$.  The second modulus $r_z$
cannot be stabilized by the same scalar field. It is intriguing to
consider leaving it unstabilized, especially since the corresponding
warping is minor, and a slow temporal variation would have very
interesting cosmological ramifications.  However, $r_z$ can also be
stabilized by a six-dimensional analogue of the GW mechanism, as we
have demonstrated here. Although the form of the effective potential
for $r_z$ is much more complicated than that in the minimal RS
scenario, a numerical analysis shows that a minimum does exist and
reproduces the desired hierarchy without the need for any fine
tuning. Indeed, the phenomenologically acceptable domain in the
parameter space of the theory is more extensive than that in the RS
model.  As for the back-reaction, while it can be incorporated,
closed-form analytic solutions are not possible owing to the non-zero
induced cosmological constants on the constant--$x_5$
hypersurfaces. However, numerical solutions are indeed possible.

In the other regime of the theory, characterized by a vanishingly
small induced cosmological constant, the scenario changes
dramatically, with the bulk tending to become conformally flat (and
the warp factor nearly exponential). With the induced cosmological
constant on the branes being infinitesimally small\footnote{Note that
a non-zero value for the five-dimensional cosmological constant 
does not preclude a vanishing four-dimensional cosmological constant 
(witness the original RS model), and, indeed, we do obtain the latter 
even in the general case. Furthermore, a vanishing value 
of the former is not a requirement for our analysis, and serves only to 
simplify the algebra.}, a closed form
solution can be found even on the inclusion of the
backreaction. This allows us to stabilize $r_z$ without taking
recourse to any unnatural values of the parameters. And while the
aforementioned exact solution is achievable for only certain specific
potentials, deviations thereof still lead to stabilization (with
backreaction taken into account) with the only difference being that
the solutions can be expressed only in terms of complicated integrals.

The situation with the corresponding $R_y$ is more intriguing.  With
$c$ now being infinitesimally small, it is tempting to consider the
possibility of a rolling $R_y$, especially since a slowly varying
$R_y$ would have very interesting and attractive cosmological
consequences. On the other hand, $R_y$ can indeed be stabilized, by
introducing a second scalar on one of the two 4-branes, viz. at $x_5 =
\pi$ or at $x_5 = 0$. The first alternative naturally leads to $R_y$
being stabilized to a value of the order of $r_z$. The second
alternative, on the other hand, leads to a situation whereby a scalar
field of apparently TeV-range mass (on account of the warping) leads
to $R_y$ being stabilized at a scale somewhat higher than the
electroweak one. If the SM fields were considered to be
five-dimensional ones, defined on the entire brane at $x_5 = 0$, this
immediately leads to a UED-like scenario with the TeV-scale protected
naturally. The orbifolding inherent to the system would not only
eliminate unwanted modes, but also introduce for a KK-parity that, in
turn, provides for a Dark Matter candidate on the one hand and
eliminates many contributions to rare decays and precision variables
on the other, thereby improving agreement with observed
phenomenology. It might be argued, though, that with $c$ being
different from zero, the KK-parity is not exact.  This is indeed so,
but with the extent of $Z_2$-breaking being determined by the
(vanishingly small) induced cosmological constant $\tilde{\Omega}$,
the lifetime of such DM-candidates would be exceedingly long.

Before ending, we revisit the question of fine tuning in such
  models. Both the original formulation~\cite{Choudhury:2006nj} as
  well as the extended version (Sec.\ref{6warped}) seemed to be
  dependent on the presence of particular values of brane tensions.
  Exactly analogous to the original RS model, this could be
  interpreted as a fine-tuning endemic to this class of
  models. Naively, the choice of values for $b_2$ and the constant of
  separation $\widetilde \Omega$ represent additional fine-tunings. We
  have shown here, though, this is not so. The exact value of
  $\widetilde \Omega$ (equivalently, $b_2$) has relatively little
  bearing on the phenomenology. Indeed, for any two values of $b_2$,
  in the range $0 \leq b_2 < \infty$, the difference between the
  resultant warp factors differs by at most $50\%$, and that too, only
  for small $k$. For large $k$, on the other hand, the warp factors
  are virtually indistinguishable except for very close to the IR
  brane. With the physical observables being differentiable functions of 
  $b_2$ (and, hence, $\widetilde \Omega$), the (small) differences due to 
  finite values of $b_2$ can be easily worked out by interpolating between 
  the results for $b_2 = 0$ and $b_2 \rightarrow \infty$ respectively. We have, 
  consequently, chosen to demonstrate the results in these two limits as 
  they admit simple analytical solutions whereas the general $b_2$ would 
  need numerical methods to be employed.

Having argued that a specific value of $\widetilde \Omega$ (or,
  equivalently, $b_2$) does not imply any fine-tuning over and above
  that endemic to RS models, we now turn to the latter, or more
  specifically, to the analogue thereof. As we have argued earlier,
  within the original RS model (sans modulus stabilization), the brane
  tension had to be just so, for the bulk solution and the orbifolding
  to be valid simultaneously. Furthermore, these were not related to
  the modulus.  Here too, a similar situation seems to hold (see
  eq.\ref{brane_potential_5}), with the recognition that the ratio
  $k/r_z$ is determined entirely by fundamental scale $M_6$ and the
  bulk cosmological constant $\Lambda_6$. Indeed, it has parallels
  with the RS model wherein the branes were allowed to have a
  nonvanishing cosmological constant. On introducing the stabilization
  mechanism the brane tensions $V_{3,4}$ were identified with the
  stabilized values of the brane-localized potentials $U_{1,2}(\phi)$
  of the bulk scalar $\phi$.

A deeper understanding is afforded if one considers possible quantum
corrections to the bulk Einstein-Hilbert action. While no such actual
calculation is available, these, presumably, would appear as
diffeomorphism-invariant higher derivative terms. Assuming that these
could be parametrized as a polynomial in $R$, Ref.\cite{sampurn}
considers, for example, a 5-dimensional bulk theory defined in the
Jordan frame by
  \[
  f(R) = R + a_1 \, \frac{R^2}{M^2} + a_2 \, \frac{R^3}{M^4} \ ,
  \] 
with the constants $a_i \sim {\cal O}(10^{-1})$ and $M$ the cut-off
scale. In the Einstein frame (obtained from the Jordan frame 
  through a conformal transformation), the extra degree of freedom 
associated with the higher-derivatives can be recast in
terms of a scalar field with a very nontrivial potential. Most
  interestingly, this degree of freedom can play the role of the
  Goldberger-Wise scalar, thereby allowing for a ``geometric 
stabilization'' of the
  modulus. A similar stabilization can occur in six-dimensions
  too with the field $\phi_1$ parameterizing such higher derivative 
    terms appearing in the bulk action.
Additional possibilities arise in the shape of brane-localized
$f(R)$-terms (since the branes are characterized by matter fields, the
quantum corrections to the Einstein-Hilbert action would, in general,
be different on them). These extra terms would play the role of the
brane-localized fields (masquerading as our $\phi_2$) with their own
potentials. Being very steep~\cite{sampurn}, these
would enforce the system being in the vacuum state, thereby according
a quantum origin to $V_{3,4}$. Of course, once again, this entire
paradigm depends upon the exact form of $f(R)$, including the
coefficients, for both the bulk and the branes. However, the
conjecture that the entire stabilization process is but a consequence
of an effective geometric action born of quantum corrections, is,
undoubtedly, a very interesting one, especially in the quest 
to understand the fine-tuning problem (such as that associated with 
choosing $\Omega = 0$).

It is also worthwhile to consider extending the formalism developed
herein to still higher dimensions. For example, it has been
shown~\cite{Dobrescu:2001ae, Burdman:2006gy, Appelquist:2001mj} that a
six-dimensional UED model not only suppresses proton decay through a
higher dimensional operator, but also gives a topological origin for
the number of chiral fermion generations. The extension of the
formalism presented here to seven dimensional nested
warping~\cite{Choudhury:2006nj} would accord a dynamical origin to the
scale of the model. These and other issues are currently under
investigation.


\section*{Acknowledgements}
We are grateful to Soumitra SenGupta, Joydip Mitra and Ashmita Das for
valuable discussions and comments.  MTA would like to thank UGC-CSIR,
India for assistance under Senior Research Fellowship Grant
Sch/SRF/AA/139/F-123/2011-12.  DC acknowledges partial support from
the European Union’s Horizon 2020 research and innovation programme
under Marie Sklodowska-Curie grant No 674896, and the R\&D grant of
the University of Delhi.

\end{document}